\documentclass[12pt,twoside]{article}
\usepackage{graphicx}

\begin{document}

\centerline{\bf The internal-environment model of the
Stern-Gerlach experiment}

\bigskip

\centerline{M. Dugi\' c$^1$, M. Arsenijevi\' c$^1$, J. Jekni\'
c-Dugi\' c$^2$}

\bigskip

\centerline{\it $^1$Department of Physics, Faculty of Science,
Kragujevac, Serbia}

\centerline{\it $^2$Department of Physics, Faculty of Science,
Ni\v s, Serbia}

\bigskip

\begin{flushleft}



\end{flushleft}

\bigskip

{\bf Abstract:} The standard interpretation of the Stern-Gerlach
experiment assumes that the atomic center-of-mass plays the role
of ``quantum apparatus'' for the atomic spin. Following a recent,
decoherence-based, model fitting with this interpretation, we
investigate whether or not such model can be constructed. Our
conclusions are somewhat surprising: only if the screen capturing
the atoms in the experiment brings the information about the
atomic-nucleus center-of-mass, one may construct the model
desired. The nucleus $CM$ system is monitored by the nucleus
``relative system ($R$)''. There appear the effective (the
electrons-mediated) interaction between $CM$ and $R$ that is
possibly responsible for decoherence. For larger atoms, the
interaction scales as $Z^2$ ($Z$ is the ``atomic number''), being
totally independent on the atomic mass. The interaction selects
the $CM$-wave-packet states as the approximate pointer basis.
Interestingly enough, the model stems nonoccurrence of
decoherence due to the internal environment for the larger
systems (such as the macromolecules and the macroscopic systems).
Certainly, disproving  this model (e.g. in an experiment) stems
the active role of the screen, which becomes responsible for the
``$CM$ + spin''-state ``reduction'' (``collapse''), i.e. for the
irreversible retrieval of the classical information from the
quantum world.

\bigskip

\section{Introduction}

\noindent The Stern-Gerlach experiment is a paradigm of the
quantum measurement of spin $\frac{1}{2}$. The standard
interpretation of the experiment (cf. e.g. [1, 2, 3])
distinguishes the atomic center-of-mass system as playing the
role of the quantum ``apparatus'' that brings the (classical)
information about the atomic spin. The interpretation assumes the
definite trajectory of every single atom (in an ensemble of
atoms) that is in one-to-one correspondence (classical
correlation) with the spin-projection along the axis of the
external magnetic field. However, in the more refined analysis,
this interpretation calls for the additional arguments.

The point is the role of the final screen capturing the atoms in
the experiment. First, in the case where the screen is a passive
element simply recording the objectively existing trajectories
(as  distinguished above) the screen serves for the so-called
quantum measurement of the second kind (the ``retrospective''
measurement) [3]-the screen does not bring any decision, it just
records the objectively present trajectories. On the other side,
having the ``active role'' in the experiment, the screen provides
the information about the trajectories, which have never
objectively existed before the screen. In the terms of the von
Neumann's quantum measurement theory [3]: the screen plays an
active role in the ``reduction'' (``collapse'') of the
center-of-mass state.

Formally, the first scenario calls for the ``mixed'' state of the
``center-of-mass+spin (CM+S)'' composite system \emph {in front}
of the screen, e.g.:
\begin{equation}
\hat\rho_{CM+S}=\frac{1}{2}\vert\uparrow\rangle_{CM}\langle\uparrow\vert\otimes\vert{-}\rangle_{S}\langle{-}\vert
+
\frac{1}{2}\vert\downarrow\rangle_{CM}\langle\downarrow\vert\otimes\vert{+}\rangle_{S}\langle{+}\vert
\end{equation}

\noindent where $\vert\uparrow\rangle_{CM}$ and
$\vert\downarrow\rangle_{CM}$ are the states of the
``center-of-mass'' system (up and down trajectories, respectively)
and $\vert{-}\rangle_{S}$ and $\vert{+}\rangle_{S}$ are the
eigenstates of the spin projection along the axis of the external
field. The second scenario, however, assumes the pure, entangled
state in the composite system \emph{in front} of the screen e.g.
of the simplified form (but see [4] for more details):
\begin{equation}
\vert\Psi\rangle_{CM+S}=\frac{1}{\sqrt{2}}\left(\vert\uparrow\rangle_{CM}\vert{-}\rangle_{S}
+\vert\downarrow\rangle_{CM}\vert{+}\rangle_{S}\right).
\end{equation}

Interestingly enough, there does not seem to appear a room for
another interpretation of the experiment: the trajectories are
either objectively present in front of  the screen, or they are
not. In the terms of the quantum measurement theory [3], one may
say that the quantum-state ``reduction'' (``collapse'') occurs--if
at all--\emph{either} in-front-of \emph{or} just on the screen.

Deciding which interpretation is correct is a subtle task,
indeed. To this end, recently, a qualitative model (yet, in
principle, allowing experimental  test) has been proposed [5].
Actually, following the first scenario (stemming yet from the
traditional wisdom), a need for the decoherence-like process for
the $CM$ system has been elevated. Namely, the mixed state eq.
(1) requires the decoherence-like process [6, 7, 8], which might
be responsible for the appearance of the ``objective''
trajectories in front of the screen. Interestingly enough, in the
context of this (standard) interpretation, it seems that the
occurrence of decoherence does not have any alternative.

Following this finding, the so-called ``relative coordinates''
(the subsystem $R$) of an atom  has  been proposed to play the
role of the ``environment'', which might  be responsible [5-8]
for the appearance of the trajectories of the atom in front of the
screen. This qualitative picture offers  a consistent
interpretation of the Stern-Gerlach experiment. First, it
completes the standard (and widely used) interpretation of the
experiment, and, second it implies non-existence  of the definite
trajectories (i.e. implies appearance of the spatial interference)
for the quantum particles not bearing any internal structure (the
subsystem $R$), such as e.g. the free electrons.

  In this paper, we  elaborate the model of  Ref. [5]. Actually, we start an ``{\it ab  initio}''
  analysis of the atomic system {\it in search for the interaction} between
  $CM$ and $R$, denoted as $\hat{H}_{CM+R}$, that might be responsible for the decoherence [5, 7,
  8], i.e. for the
  appearance of the $CM$'s ``objective'' trajectories in front of the screen.
  To this end, we find that ``$CM$'' is the {\it nuclear-system} center-of-mass,
  not yet the center of mass of the atom as  a whole (in contrast to the wide-spread believing [1, 2, 3]),
  while the corresponding interaction is a reminiscent of the Coulomb interaction between the electrons
  and the protons in the atom. Our quantitative estimations stem
  the relative increase of the strength of $\hat H_{CM+R}$ for the
  larger atoms. Interestingly enough, the strength scales as $Z^2$
  ($Z$ being the so called  atomic number) for  larger atoms, being yet totally independent on the mass
  number $A$. In other words: as distinct from the traditional expectation, the atomic
  mass does not appear as a parameter of importance for the
  interpretation of the SG experiment, i.e. for the occurrence of decoherence in $CM$ system.

          Of course, in the case of the physical (e.g. the experimental) disproving of our conclusions
          (i.e. rejecting the decoherence-based  model), there remains the alternative scenario--cf.
          eq. (2)--as the only remaining candidate for interpretation of the Stern-Gerlach experiment.
          Needless to say, this possibility might reinforce the interpretation of the non-repeatable
           quantum measurements as the basic source of the irreversible retrieval
          of the classical information from the quantum world.

\section {The atomic subsystems}

\noindent``Atom'' is a complex system--it consists of the
electrons, the protons and in general the neutrons. This
fundamental  definition is however usually inaccessible
experimentally. Actually, likewise for the macroscopic systems,
the atoms
 (but also the molecules)
 are usually observed by observing their center-of-mass degrees of freedom.
 This is actually the case in the SG experiment: the screen captures the
 atoms thus revealing their center-of-mass positions, while usually not providing us
 with any information about the internal state; this is the reason one may assume
 the ground internal-state of the atom.

In general, the  linear canonical transformations
\begin{equation}
\hat{\vec{R}}_{_{CM}}={\displaystyle\sum_{i=1}^N
m_i\hat{\vec{r}}_i}/{\displaystyle\sum_{i=1}^N m_i}
\end{equation}
introduce the center-of-mass ($CM$) of an $N$-particle
system, likewise  the ``relative coordinates''($R$)
\begin{equation}
\hat{\vec{\rho}}_{R\alpha}=\hat{\vec{r}}_i-\hat{\vec{r}}_j
\end{equation}
where $\hat{\vec{r}}_m$ represents  the position of the  $m$-th
constituent particle of the composite system. The inverse
transformations also hold
\begin{equation}
\hat{\vec{r}}_i=\hat{\vec{R}}_{CM}+\displaystyle\sum_{\alpha=1}^{N-1}\omega_{\alpha}\hat{\vec{\rho}}_{{R}\alpha}\!\!\!^{\scriptstyle{(i)}},
\end{equation}
where  $\omega_{\alpha}=m_{\alpha}/{M}$ ($m_{\alpha}$'s are the
masses of the constituent particles, while $M$ is the total mass
of  the system). Bearing in mind eqs. (3), (4), (5), we point out
the first observation as made in [5]: introducing the $CM$
 system necessarily calls for the ``relative system'' $R$ as formally described
(non-uniquely) by the relative coordinates eq. (4). The standard
model of the Stern-Gerlach experiment fully discards the
(sub)system $R$ from the considerations, while assuming $R$ to be
in the ground state, and the atomic spin being describable
solely  by the state of the valent  electron. And so appears the
basic idea of the decoherence-based approach [5]: introducing the
system $R$ as the $CM$'s environment might complete the physical
picture as described in Introduction.

\subsection{The general task}

\noindent We search for the possible physical origin of the
interaction $\hat{H}_{{CM+R}}$,
 which is necessary [6, 7, 8] for the occurrence of the $CM$'s decoherence.
More precisely: our starting point is the fundamental
 model of an atom--defined as a collection of the electrons, protons
 and neutrons--and we search for the (effective) interaction between
 the properly defined $CM$ and $R$ systems.
 Once having a proper model of interaction, we investigate the possible
  ``pointer basis (states)'' and the relative strength of the interaction
as  essential for  the description of the desired decoherence
effect. The analysis bears some subtlety yet, as apparent in the
sequel.

In order to elaborate this  decoherence--based model, we begin
from the very definition of the atomic Hamiltonian, for the atom
defined as a collection of the  electrons, protons and the
neutrons. Then, the  atomic Hamiltonian reads:
\begin{equation}
\hat{H}=\sum_{i=1}^Z\hat{T}_{Ei}+\sum_{j=1}^Z\hat{T}_{pj}+\sum_{k=1}^{A-Z}\hat{T}_{nk}+\hat{V}^{ee}_{Coul}+
\hat{V}^{ep}_{Coul}+\hat{V}^{pp}_{Coul}+\hat{V}_{nucl}
\end{equation}
with the  following  meaning for each term: $\hat{T}$ stands for
the kinetic terms, $\hat{V}_{Coul}$ for the Coulomb interaction
 of the pairs of particles ({\it ee}-the electrons, {\it ep}-the electron-proton,
{\it pp}-the protons pairs), and the nucleon interaction for a
pair $(n, n')$ of nucleons is given e.g. by [9]:
\begin{equation}
\hat{V}_{nucl}^{nn'}\equiv{-}\gamma^{2}\displaystyle\frac{\exp({-\mu|\hat{\vec{r}}_{n}-\hat{\vec{r}}_{n{'}}|})}{|\hat{\vec{r}}_{n}-\hat{\vec{r}}_{n{'}}|}
\end{equation}

\noindent where $\gamma$ is a constant and  r=$\frac{1}{\mu}$ is
the range of the nuclear forces. For simplicity, we omit the
comparatively weak interactions, such as the spin-spin or
spin-orbit interactions in the atom.

\bigskip

\noindent {\it (a) The atomic center of mass}

\bigskip

\noindent We apply the linear canonical transformations eqs. (3),
(4) to the fundamental model of ``atom'', eqs. (6), (7). Instead
of the model ``electrons + protons + neutrons'', we introduce the
center of mass $CM$ and the ``relative system'' $R$ of the atom as
a whole, thus defining the following tensor-product structure of
the atomic Hilbert state space: $\mathcal{H} = \mathcal{H}_{CM}
\otimes \mathcal{H}_R \otimes \mathcal{H}_S$, where the index $S$
refers to the atomic spin.

Bearing in mind the standard model of the SG experiment, the atom
exposed to the external magnetic field is described by the
following Hamiltonian (that is a straightforward transform of eq.
(6)):
\begin{equation}
\hat H = \hat T_{CM} + \hat H_R + \hat H_{CM+S},
\end{equation}

\noindent where $\hat H_{CM+S} = \mu_B B(\hat z_{CM}) \otimes
\hat S_z$ is the standard term [5] coupling the external magnetic
field (for simplicity lying along the $z$-axis) and the atomic
spin $\hat S_z$, while the $R$-system's self-Hamiltonian reads:
\begin{equation}
\hat H_R
=\sum_{\alpha=1}^{Z+A-1}\hat{T}_{R\alpha}+\hat{V}^{(R)}_{nucl}+\hat{V}^{(R)}_{Coul}+\hat{M}^{(R)}_{\eta\nu},
\end{equation}
in analogy with eq. (8), and  $\hat{M}^{(R)}_{\eta\nu}$ is
the internal interaction (cf. Appendix 1).

Regarding eq. (9), it is important to note: being the
distance-dependent, all the original interactions (the Coulomb
interactions and the nuclear interactions  in eq. (6)) transform
into the ``external fields'' (the one-particle potentials $V(\hat
{\vec \rho}_{Ri})$) for the ``relative particles''  system
 $R$ (cf. Appendix 1). These effective potentials are the terms of the $R$'s self-Hamiltonian $\hat H_R$.
 Needless to say, this gives the exact separation of $CM$ and $R$ that does  not leave a room for the desired decoherence of the $CM$ states.

In the terms of the quantum states, the initial state, e.g.,
$$|\Psi \rangle_{CM}|0 \rangle_{R}\frac{1}{\sqrt{2}}(|+\rangle_S+|-\rangle_S)$$
(where we omit the symbol of the tensor product) dynamically
transforms as presented by the following simplified expression:
\begin{equation}
\hat U |\Psi \rangle_{CM}|0 \rangle_{R}
\frac{1}{\sqrt{2}}(|+\rangle_S+|-\rangle_S)
=\frac{1}{\sqrt{2}}(|\uparrow\rangle_{CM}|0\rangle_{R}|-\rangle_S+
|\downarrow\rangle_{CM}|0\rangle_{R}|+\rangle_S).
\end{equation}

\noindent $\hat U$ is the unitary operator of evolution in
time generated by the Hamiltonian eq. (8)--recall that the system $CM + R + S$ (plus the external {\it
classical} magnetic field) is an isolated quantum system. By
tracing out the system $R$, one obtains the entangled state eq. (2).

So, we conclude: the introduction of $CM$ and $R$  of the atom as
a {\it whole} does not provide the occurrence of decoherence as
desired in Section 1. Interestingly enough, this is not the end
of the program, as it is shown in the sequel.

\bigskip

\noindent{\it (b) The nucleus center of mass}

\bigskip

\noindent More than 99.99 per-cents of the atomic mass is placed
in the atomic nucleus. Practically, it is truly hard to
distinguish between the atomic and the nucleus center-of-mass
systems. So, we investigate another application of eqs. (3), (4):
we introduce the center-of-mass system and the ``relative
system'' for the atomic {\it nucleus} while leaving the electrons
variables intact.

Introducing the collective degrees of freedom of the atomic
nucleus is the standard procedure in nuclear physics [10, 11]. On
the other side, the similar idea appears in certain models of the
quantum measurement theory, unfortunately not yet being fully
elaborated [12]. So, introducing the center of mass of the atomic
nucleus not yet involving the electrons is physically legitimate
a procedure. Its relation to the procedure presented in (a) will
be discussed in Section 4.

Then, ``atom'' is a composite system defined as ``$E + CM + R +
S$'', where $E$ stands for the electrons-system, $CM$ and $R$ for
the nucleus center-of-mass and the ``relative'' systems,
respectively,  while $S$ is the atomic spin. The corresponding
Hilbert state space of ``atom'' now reads: $\mathcal{H} =
\mathcal{H}_E \otimes \mathcal{H}_{CM} \otimes \mathcal{H}_R
\otimes \mathcal{H}_S$; we hope the use of the same notation for
the center-of-mass and the relative system as in (a) will not
produce any confusion--further on, we are only concerned with the
case (b) under consideration.

Now, the standard model of the SG experiment is defined by the
following form of the atomic Hamiltonian (in analogy with eq.
(8)):
\begin{equation}
\hat H = \hat H_E + \hat T_{CM} + \hat H_R + \hat H_{CM+S} + \hat
H_{E+CM+R}.
\end{equation}

\noindent Certainly, the Hamiltonian $\hat H$ in eq. (8) and eq.
(11) is the one and the same observable--it is just written in
the different forms, yet in eq. (11) appearing the interaction
term for $E$, $CM$ and $R$ systems (cf. Appendix 1):
\begin{equation}
\hat{H}_{E+CM+R}=k\displaystyle\sum_{i=1}^Z
\displaystyle\sum_{j=1}^Z\displaystyle\frac{1}{|\hat{\vec{r}}_{Ei}-\hat{\vec{R}}_{CM}-\displaystyle
\sum_{\alpha=1}^{A-1}\omega^{(j)}_{\alpha}{\hat{\vec{\rho}}}_{R\alpha}\!\!\!^{\scriptstyle{(j)}}|}.
\end{equation}
Physically, the tripartite interaction $\hat H_{E+CM+R}$ is a
particular form of the Coulomb interaction between the atomic
electrons and the protons. Fortunately enough, this tripartite
interaction can be reduced to a bipartite interaction coupling
$CM$ and $R$ systems--cf. below.

The close inspection of the rhs of eq. (11) justifies the
 application of the {\it adiabatic approximation} that
in its zeroth order separates the electrons system from the rest
(cf. Appendix 2). More precisely: the electrons are too light
relative to both the $CM$- and $R$-mass, thus allowing the
standard procedure of the adiabatic approximation [2, 13, 14]. On
the
 other side, for the realistic atoms (Appendix 2), the $CM$ and $R$
mass-ratio does not allow the application of the adiabatic
 approximation. So, we expect the approximate separation of
the electrons-state from the rest, $CM + R + S$,  as well as
non-negligible entanglement between $CM$ and $R$. Formally,
 the state of this decomposition of ``atom'' in subsystems reads:
\begin{equation}
\vert \chi \rangle_E \vert \Phi\rangle_{CM+R+S} + \vert
O(\kappa)\rangle_{E+CM+R+S},
\end{equation}

\noindent where the small term (that bears entanglement, in
general, of all of the subsystems) is of the norm $\kappa^{3/4}$,
where $\kappa = max \{\kappa_1, \kappa_2\}$, and $\kappa_i$ are
the corresponding mass ratios, cf. Appendix 2.

In order to obtain the dynamics of the "slow" system $CM + R$
(i.e. of $CM + R + S$), one should discard the electrons system
as (cf. Appendix 2):

\begin{equation}
\hat H_{CM+R+S} \equiv _E\langle \chi \vert \hat H \vert \chi
\rangle_E \cong \hat T_{CM} + \hat H_R + \hat H_{CM+S} + \hat
H_{CM+R},
\end{equation}

\noindent where
\begin{equation}
\hat H_{CM+R} \equiv _E\langle \chi \vert \hat H_{E+CM+R} \vert
\chi\rangle_E
\end{equation}
represents the {\it effective} (the electrons--system {\it mediated})
interaction between $CM$ and $R$.

Now, due to the two interaction terms, $\hat H_{CM+S}$ and $\hat
H_{CM+R}$ in eq. (14), it is straightforward dynamically to
obtain entanglement in the dominant term of the state in eq.
(13), $\vert \Phi\rangle_{CM+R+S}$. Actually, for the initial
state $$|\Psi \rangle_{CM}|0 \rangle_{R}\frac{1}{\sqrt{2}}(|+\rangle_S+|-\rangle_S),$$
it is easy to obtain entanglement (compare to eq. (10)) as:
\begin{equation}
\hat U |\Psi \rangle_{CM}|0 \rangle_{R}
\frac{1}{\sqrt{2}}(|+\rangle_S+|-\rangle_S)
\cong\frac{1}{\sqrt{2}}(|\uparrow\rangle_{CM}|1\rangle_{R}|-\rangle_S+
|\downarrow\rangle_{CM}|2\rangle_{R}|+\rangle_S).
\end{equation} where $\hat{U}$ is generated by $\hat{H}$
represented in  eq. (14). Now,
assuming the orthogonality $_R\langle1|2\rangle_R\approx0$, by tracing out the
``environment'' $R$ from the  rhs of eq. (16) follows the
mixed state eq. (1) for $CM+S$ system.

\bigskip
{\bf 3. The occurrence of decoherence in SG experiment}

\bigskip

\noindent In general, the interaction between the open system and
its environment is a necessary, not yet the sufficient condition
for decoherence. The occurrence of decoherence is a subtle
 an issue, indeed [7, 8], formally depending on the (at least approximate) orthogonality of
 the environmental states--e.g., the states $|i\rangle_R$ $(i=1,2)$ on the rhs of eq. (16).
 So, the presence of the interaction $\hat H_{CM+R}$ opens in principle a room for decoherence
 of $CM$'s states that are yet to be determined.

The occurrence of decoherence generated by $\hat H_{CM+R}$
competes with the {\it recurrence} that bears the two origins.
First, the dynamics generated by the open system's
self-Hamiltonian ($\hat H_{CM}$) produces coherence as--cf.
below--not commuting with the interaction term ($\hat H_{CM+R}$).
Second, the overall dynamics (generated by the total Hamiltonian
$\hat H$) is unitary, and for the small environment, the
recurrence of the initial coherence may be effective.
Unfortunately, the method of the unitary operator $\hat U$
applied here does not in general allow the estimations of the
decoherence time and of the recurrence time. On the other side,
by investigating solely the interaction term, one may conclude
about the possible ``pointer basis'' of the open system $CM$ [7,
8].

Therefore, the task for this section reads: to calculate the
interaction term $\hat H_{CM+R}$ and to estimate both the
possible ``pointer basis'' states (within the context of the
standard model of SG experiment) as well as the dependence of the
interaction on the size of the atom; following eq. (12), the
``size of the atom'' is determined by the number of the electrons
(protons), $Z$, not yet by the ``mass  number'' $A$.
This way, we estimate the relative strength of the
interaction and its efficiency in producing decoherence for the $CM$'s
pointer basis states. Investigating the decoherence- and the
recurrence-time requires the different methods (e.g. certain types of the master
equations, cf. e.g. [15]) and comes out of the scope of the
present paper.

\bigskip

{\bf 3.1 The interaction $\hat{H}_{CM+R}$}
\bigskip

\noindent The dynamics of the slow system $CM+R+S$ is generated
by  $\hat{H}_{CM+R+S}$, eq. (14). However, essential for our
program is the interaction term eq. (15). Substitution of eq. (12)
in eq. (15) directly gives:
\begin{equation}
\hat{H}_{CM+R}=k{\displaystyle\sum_{i=1}^Z}{\displaystyle\sum_{j=1}^Z}{_E\langle\chi|\displaystyle\frac{1}{|\hat{\vec{r}}_{Ei}-\hat{\vec{R}}_{CM}-\displaystyle\sum_{\alpha=1}^{A-1}\omega^{(j)}_{\alpha}\hat{\vec{\rho}}_{R\alpha}\!\!\!^{\scriptstyle{(j)}}|}|\chi\rangle_E}
\end{equation}
where $|\chi\rangle_E$ are the eigenstates of the adiabatically-defined
electrons-system self-Hamiltonian $\hat{H}_E$. Being
interested in the $Z$-dependence of $\hat{H}_{CM+R}$, we introduce
the following choice: for $|\chi\rangle_E$, we take the
$Z$-electrons Slater determinant constructed from the hydrogen-atom
states. So, we neglect both, the adiabatic approximation as well
as the corrections for the many-electrons atoms. We believe that
these simplifications will not significantly  change the result,
while bearing in mind yet that the states appearing in
$|\chi\rangle_E$  refer to the point-like nucleus.

The expression eq. (17) bears  formally identical terms for all
the protons, so we can write:
\begin{equation}
\hat{H}_{CM+R}=Z\hat{H}^{'}_{CM+R}
\end{equation}
where
\begin{equation}
\hat{H}^{'}_{CM+R}=k{\displaystyle\sum_{i=1}^Z}{_E\langle\chi|\displaystyle\frac{1}{|\hat{\vec{r}}_{Ei}-\hat{\vec{R}}_{CM}-\displaystyle\sum_{\alpha=1}^{A-1}\omega_{\alpha}{\hat{\vec{\rho}}}_{R\alpha}|}|\chi\rangle_E},\forall
j.
\end{equation}

Due to the exact orthogonality of the one-electron states
$|\phi_i\rangle_E$ appearing in $|\chi\rangle_E$, one easily
obtains:
\begin{equation}
\hat{H}^{'}_{CM+R}=k{\displaystyle\sum_{i=1}^Z}{_E\langle\phi_i|\displaystyle\frac{1}{|\hat{\vec{r}}_{Ei}-\hat{\vec{R}}_{CM}-\displaystyle\sum_{\alpha=1}^{A-1}\omega_{\alpha}{\hat{\vec{\rho}}}_{R\alpha}|}|\phi_i\rangle_E},
\end{equation}
where the $i$ enumerates the electrons, i.e. the states
$|\phi_i\rangle_E$ appearing in the Slater determinant
$|\chi\rangle_E$.

Measuring (as usual in quantum mechanics of molecules [13, 14])
the electrons positions from the (point-like) atomic nucleus,
$\hat{\vec{r}}_{Ei}\longrightarrow\hat{\vec{r}}_{Ei}-\vec{r}_{CM}\hat{I}_E$,
the expression (20) obtains the following
form in the position-representation:
\begin{equation}
\hat{H}_{CM+R}=kZ\displaystyle\sum_{i=1}^Z\int
\displaystyle\frac{|\phi_i(\vec{r}_{Ei}-\vec{r}_{{CM}})|^2}
{|\vec{r}_{Ei}-\hat{\vec{R}}_{CM}-\displaystyle\sum_{\alpha=1}^{A-1}\omega_{\alpha}{\hat{\vec{\rho}}}_{R\alpha}|}d^{3}\vec{r}_{Ei}.
\end{equation}
By introducing the ``shift'':
$\vec{r}_{Ei}-\vec{r}_{{CM}}\equiv\vec{\xi}_i$, one obtains the
following form for $\hat{H}_{CM+R}$ (while omitting the unnecessary
index $i$ in the vector $\vec{\xi}_i$):
\begin{equation}
\hat{H}_{CM+R}=kZ\displaystyle\sum_{i=1}^Z\int\displaystyle\frac{|\phi_i(\vec{\xi})|^2}
{|{\vec{\xi}}-\hat{\vec{\Omega}}_{CM+R}|}d^{3}\vec{\xi}
\end{equation}
\noindent where
$\hat{\vec{\Omega}}_{CM+R}\equiv-\vec{r}_{{CM}}\hat{I}_E+\hat{\vec{R}}_{CM}+\displaystyle\sum_{\alpha=1}^{A-1}\omega_{\alpha}\hat{\vec{\rho}}_{R{\alpha}}$.

The method for calculating (22) can be found in Ref. [16]. The
details are presented in Appendix 3, and the result for the atoms
with the ``closed shells'', reads :
\begin{eqnarray}
\hat{H}_{CM+R}&=\nonumber&kZ\sum_{n}\sum_{\ell=0}^{n-1}\sum_{g=0}^{n-\ell-1}\sum_{t=0}^{2g}\frac{(2\ell+1)}{2n
2^{2(n-\ell-1)}}{{2(n-\ell-1)-2g}\choose{n-\ell-1-g}}\times
\nonumber \\&&
\times\frac{(2g){!}}{g{!}(2\ell+1+g){!}}{{2g+2(2\ell+1)}\choose{2g-t}}\frac{(-2)^t}{t{!}}\nonumber\\&&\Bigg\{(2\ell+t+2){!}
\Bigg(1-\exp\bigg(-\frac{2Z\hat{\Omega}}{na_{\mu}}\bigg)\sum_{f=0}^{2\ell+t+2}\frac{(\frac{2Z\hat{\Omega}}{na_{\mu}})^{f}}{f{!}}\Bigg)\hat{\Omega}^{-1}\nonumber\\&&+
\frac{2Z}{na_{\mu}}(2\ell+t+1){!}\exp\bigg(-\frac{2Z\hat{\Omega}}{na_{\mu}}\bigg)\sum_{f=0}^{2\ell+t+1}\frac{(\frac{2Z\hat{\Omega}}{na_{\mu}})^{f}}{f{!}}\Bigg\},
\end{eqnarray}
where $\hat{\Omega}\equiv|\hat{\vec{\Omega}}_{CM+R}|$ and $a_\mu$
is the  analog of the Bohr radius. In eq. (22), the index $i$
enumerates the electrons, i.e. the one-electron quantum states
determined uniquely  by the hydrogen-atom quantum numbers, ${
n,\ell}$ and ${m}$. This is the origin and the meaning of the
indices ${n,\ell}$ and ${m}$ in eq. (23).

\bigskip

{\bf 3.2 The pointer basis and the relative strength of
interaction}

\bigskip

\noindent The expression (23) is the effective interaction of
$CM$ and $R$ subsystems of the atomic nucleus. The interaction
stems directly from the couplings present in the exponential-,
polynomial- as well from the $\hat{\Omega}^{-1}$--terms.

More precisely, due to definition of $\hat{\Omega}$ (cf. above)
one directly infers the information about the possible pointer
basis [7, 8] for $CM$ system. It is rather apparent:
$\hat{H}_{CM+R}$ is exactly {\it diagonalizable} in the continuous
eigenbasis of the observable $\hat{\vec{R}}_{CM}$:
$\langle\vec{R}_{CM}|\hat{H}_{CM+R}|\vec{R}_{CM}^{'}\rangle=0$ for
$\vec{R}_{CM}\neq\vec{R}_{CM}^{'}$, {\it thus promoting} [7, 8]
the basis $|\vec{R}_{CM}\rangle$ as the {\it possible pointer
basis}. However, there is an even better choice of the pointer
basis, i.e. of the ``preferred set'' of (non-orthogonal) yet
normalizable states.

Actually, the minimal--uncertainty  (the ``coherent'') states
$|\Psi_{\mu\nu}\rangle_{CM}$ modeling the $CM$ wave-packets {\it
approximately diagonalize} $\hat{H}_{CM+R}$--as presented in [3,
17]. These states approximate both $\hat{\vec{R}}_{CM}$ and its
conjugate momentum-operator $\hat{\vec{P}}_{CM}$, thus allowing
one to define the semi-classical trajectories of the wave packet
modeled by $|\Psi_{\mu\nu}\rangle_{CM}$. The transition from the
exact (non--normalizable) states $|\vec{R}_{CM}\rangle$ (or
$|\vec{P}_{CM}\rangle$) to the ``coherent states''
$|\Psi_{\mu\nu}\rangle_{CM}$ is generally established by von
Neumann [3] (cf. ch. 5 therein). So, we may promote the
``coherent states'' $|\Psi_{\mu\nu}\rangle_{CM}$
 as the {\it approximate pointer basis} that, fortunately enough, fits
 with the wave-packet model [1, 2, 3, 5] of the atoms in the SG experiment.

Finally to this section, our aim is to determine the relative
strength  of $\hat{H}_{CM+R}$, i.e. to determine its
$Z$--dependence. This is certainly essential for our program: the
cases
 $Z=0,1$ can not be described by the model (b), so we want to estimate the strength
of $\hat{H}_{CM+R}$ for the different values of $Z>1$.

To this end, we first want to stress: in eq. (23) appear the
multiplications
 of the exponential functions of  $Z$ with the polynomials of $Z$. As it can be easily
 shown, already for $Z\sim10$ , one may neglect all such multiplications, thus
 obtaining the ``linear'' dependence of $\hat{H}_{CM+R}$ of $Z$ (for $Z\stackrel{\sim}{>} 10)$:
 \begin{equation}
 \Arrowvert\hat{H}_{CM+R}\Arrowvert\stackrel{\sim}{=}\beta Z\Arrowvert\hat{\Omega}^{-1}\Arrowvert
\end{equation}
where
\begin{eqnarray}
\beta&\equiv\nonumber&k\sum_{n}\sum_{\ell=0}^{n-1}\sum_{g=0}^{n-\ell-1}\sum_{t=0}^{2g}\frac{(2\ell+1)}{2n
2^{2(n-\ell-1)}}{{2(n-\ell-1)-2g}\choose{n-\ell-1-g}}\times
\nonumber\\&&
\times\frac{(2g){!}}{g{!}(2\ell+1+g){!}}{{2g+2(2\ell+1)}\choose{2g-t}}\frac{(-2)^t}{t{!}}\big[(2\ell+t+2){!}\big].
\end{eqnarray}

The appearance of the quantum numbers $n,\ell$ and $m$ in eq.
(25) apparently
 points the $Z$--dependence of the parameter $\beta$. Now, bearing in mind
 that eq. (23) refers to the atoms with the ``closed shells'', we prepared a
 numerical estimation for the $Z$--dependence of $\beta$, yet for the particular
  values: $Z$=10, 28, 60, 110, 182 (i.e. for $n$=2, 3, 4, 5, 6). The result is presented
  in Fig. 1, clearly stemming the linear dependence of $\beta$ on $Z$.

\begin{figure}[t]
  \centering
 \includegraphics{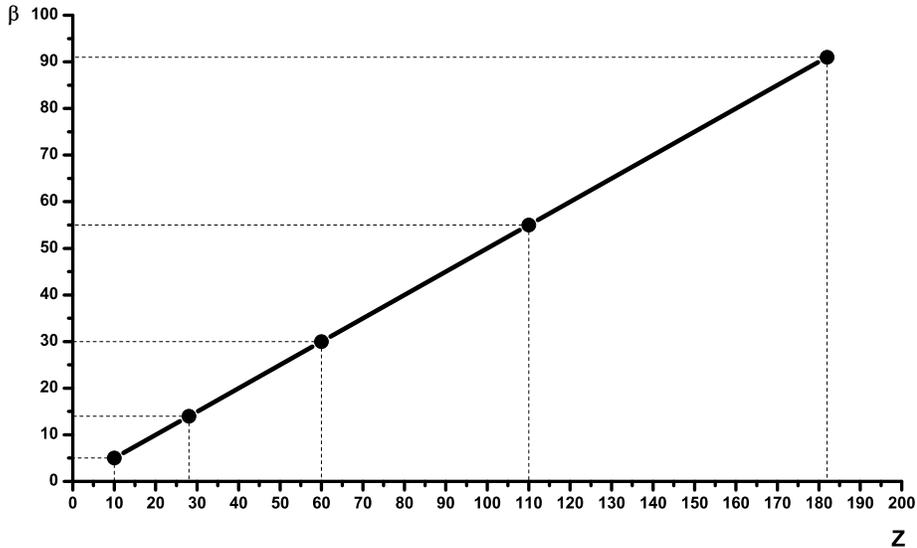}
\caption{Dependence of the factor $\beta$ on $Z$ for the atoms
with the closed shells.}
 \label{Fig. 1.}
\end{figure}

Therefore, we emphasize: except for the small atoms, i.e. for the
small values of $Z$ ($Z<10$) , one obtains  that the strength of
interaction $\hat{H}_{CM+R}$ approximately scales as $Z^{2}$,
thus clearly giving advantage to the larger atoms (with the
larger $Z$) in the sense of the possible occurrence of
decoherence, which may select (cf. above) the ``coherent states''
as the approximate pointer basis.

\bigskip

{\bf 3.3 The physical picture}
\bigskip

\noindent The possible occurrence of decoherence regarding the
model (b) gives the following physical description of the SG
experiment.

The free moving of the nucleus center-of-mass wave-packet is
adiabatically followed by the electrons. The external magnetic
field quickly induces entanglement between the $CM$ system and
the atomic effective spin $S$ [5]. The fast formation of
entanglement in $CM+S$ system can be easily shown to be followed
by the formation of entanglement between $CM$ and $R$ system (as
generated by $\hat{H}_{CM+R}$) , the later also being
adiabatically decoupled from the electrons system $E$. The
subsequent decoherence gives rise to the semi-classically
well-defined trajectories of the $CM$ system, thus giving rise to
the  mixed state eq. (1). Certainly, the screen is supposed to
serve as a passive recorder of the ``objectively'' existing
trajectories of the {\it nucleus} center-of-mass--the atomic-{\it
nucleus} center-of-mass serves as a ``quantum apparatus'' for the
atomic spin, that fits essentially with the standard
interpretation of the SG experiment.

However, as  distinct from the standard interpretation, our model
of the SG experiment removes the focus from both the {\it atomic}
center-of-mass as well as from the {\it atomic mass}. As to the
later, the parameter of interest is {\it not the atomic mass} but
the number of the protons (electrons) in the atom, $Z$. The
interaction $\hat H_{CM+R}$,  that is responsible for the
possible occurrence of decoherence,  scales (approximately) as
$Z^2$, thus clearly giving advantage to the larger atoms as
possibly the decoherent systems.

Interestingly enough, the presence of {\it both} $R$ and $E$
systems is {\it necessary} for the occurrence of decoherence.
First, nonexistence of $R$ stems nonexistence of the interaction
$\hat H_{CM+R}$. On the other side, nonexistence of the electrons
system would imply nonexistence of the tripartite
interaction-term $\hat H_{E+CM+R}$ and thus--cf. eq. (15)--again,
nonexistence of the interaction $\hat H_{CM+R}$ (in  some analogy
with the model (a)). Therefore, the model (b) stems the absence
of the SG-effect for the free electrons, protons and neutrons
(the later already being well-known from [18]), for the bare
atomic nuclei as well as for the hydrogen atom. The effect can be
considered to be rather weak for the small atoms ($Z < 10$) and
more effective for larger atoms. The precise form of $\hat
H_{CM+R}$ for the atoms with ``incomplete shells'' can be easily
obtained by adapting eq. (23)--e.g. by adding the new terms
defined by the states of the electrons not (classically speaking)
belonging to the ``closed shells''. This, however, does not seem
to change anything in our qualitative estimation of the
$Z$-dependence of $\hat H_{CM+R}$.

The physical basis of a possible experimental test of the model
(b) can be now directly borrowed from Ref. [5]. While we do {\it
not} guarantee that the {\it screen really captures the
atomic-nucleus-, instead of the atomic-center-of-mass}, the
experimentally observed absence of the effect e.g. for the
massive, bare atomic nucleus might be a significant support for
the model considered.

Needless to say, if the screen captures the center-of-mass of the
{\it atom as a whole}, one seems to be obliged to reinterpret the
SG experiment. Actually, (as emphasized in Introduction) then the
screen itself would be responsible for the decoherence-like
effect by transforming the pure state eq. (2) into incoherent
mixture eq. (1).

\bigskip

 {\bf 4. Discussion}

 \bigskip

 \noindent
 The choice of the collective degrees of freedom of atomic
  nucleus naturally--and, probably, unavoidably--calls for
  the adiabatic approximation for the atomic electrons system.
   The nuclear physics theory seems silent in this regard [11].
   The estimates of the mass-ratios (cf. Appendix 2) justify the adiabatic
   cut of the electrons from both $CM$ and $R$ systems,
   not yet justifying the adiabatic cut of $R$ from $CM$--the
   later giving
   possibly rise to the occurrence of decoherence that is
 induced by the bipartite interaction $\hat H_{CM+R}$.

The interaction model eq. (23) justifies (cf. Section 3.2) the
{\it wave-packets} as the {\it approximate} pointer basis thus
recovering the standard interpretation of SG experiment. As a
benefit, we obtain that for the larger atoms, the
 decoherence-inducing interaction is stronger, while for the free electrons
 (likewise the protons, and particularly the neutrons [18,19]),
 as well as for the hydrogen atom, one can not expect decoherence due to
 nonexistence of the subsystem $R$. Of course, we have {\it not} proved the
 occurrence of decoherence--this requires the different methods
 (e.g. in estimating the decoherence- and the recurrence-time [15]).
 Nevertheless, we have prepared all the necessary data in the case the
 decoherence may be considered to be effective, yet bearing some subtlety.

Actually, one may pose the following question. The straightforward
extrapolation of our model may seem to imply the occurrence of
decoherence for the larger systems--such as the large molecules,
and even more for the macroscopic systems--due to the internal
environment. Bearing this in mind, one may ask if our model
contradicts the experimentally verified coherence (interference)
of the large-molecules' center-of-mass states [20]. Interestingly
enough, our model directly answers this question: the $R$ system
can not be considered to induce decoherence of $CM$ system for
large quantum systems. Namely, the mass ratio (Appendix 2)
$\kappa_3 = \mu /M$ decreases with the increase of the number of
particles in the system, thus in effect allowing the adiabatic
separation between $CM$ and $R$; the larger the system, the
larger the total mass $M$, while $\mu$ remains virtually
unchanged. In the formal terms: the large system's state is
presented approximately (the larger the system the better the
approximation) by the first term in eq. (13). Therefore, we
conclude that the occurrence of decoherence for the large systems
is {\it not induced by the internal environment}--as supported by
the macromolecules decoherence experiments [21].

 Finally, we should clarify the relation between the two different
 models of ``atom'' as defined by (a) and (b) in Section 2.1. Actually,
 one may ask, after all, if
 ``atom'' is an isolated or a decoherent (open) system.
 And this is another subtle issue, indeed.

 The two models are  the legitimate physical models of the
 one and the same physical system--the ``atom''. They
 represent the different faces (facets) of  ``atom'', thus
  not necessarily providing the same or even equivalent
  informations about ``atom''. To this end, a particular
  approach has recently been proposed in the context of
  the problem ``what is system'' [22-24]. Actually, the different
  models of a composite system may still give rise to the
  mutually non-equivalent physical and information-theoretic
  descriptions of the composite system itself. On the intuitive
  level, answering what is system may bear some arbitrariness,
  thus promoting {\it relativity} of the very fundamental physical
  concept of ``system'' [22, 23]. In a sense, the two models
  (a) and (b) are paradigmatic for the system-relativity issue:
  both models describe the one and the same composite system--the ``atom''--
  while the different separations of the system into subsystems
  may still give rise to the different physical descriptions of
  ``atom''.
 Now, the observation of the {\it atomic} center-of-mass would describe
 the ``atom'' as an {\it isolated} quantum system. On the other
 side--if we trust the model (b)--the observation of the {\it nucleus}
 center-of-mass (as it  might be the case in SG experiment)
 might give rise to the opposite answer--the ``atom'' may appear  as a decoherent,
 and certainly as an  {\it open} system. And the system relativity stems [23, 24]: the two
 answers do not raise a contradiction; they just represent the two
 different descriptions of ``atom'' still depending on the ``point of view'',
 i.e. on the {\it observables targeted} in the course of observation. So, there
 is not a unique answer to the above question regarding ``atom'' as
 isolated/open system. Observation of ``atom'' as isolated quantum system
 might be equally physically realistic as the observation of ``atom'' as open quantum system.

Needless to say, we always observe (or perceive) only a
``fraction'' (the comparatively small subsystems) of a composite
system, and still try to generate the proper description of the
composite system as a whole. This, however is a usual conjecture
not necessarily supported by the system-relativity [22-24]. To
this end, now appears the following question: whether or not the
models (a) and (b) may be reduced to each other, or to another
(not yet known), more fundamental model of ``atom''? In this
regard, we just want to emphasize: mutual {\it irreducibility} of
the two models seems to stem as a corollary of the standard
interpretation of the decoherence process [6]. {\it Conversely},
the
 mutual {\it reducibility of the models} (as well as existence of
 the alternative, more fundamental model) might {\it hide a clue}
 for giving the answer to the problem of the ``transition from
 quantum to classical'' [6]--e.g. to the transition from the
 isolated to the open ``atom''. In this sense, our study may
 sharpen this truly fundamental issue of modern quantum mechanics.

So, one may say, that we offer a consistent physical model  of
``atom'' that might support the standard interpretation of the
Stern-Gerlach experiment. Needless to say, in the case of e.g.
experimental disproving of this (the decoherence--based model),
there is the alternative (cf. Introduction) interpretation of the
Stern-Gerlach experiment that still requires elaboration in the
context of the non-repeatable quantum measurement which is out of
the scope of the present study.

\bigskip

{\bf 5. Conclusion}

\bigskip

\noindent If the screen in the Stern-Gerlach experiment capturing
the atoms brings the information about the atomic-nucleus
center-of-mass ($CM$), then there exists a decoherence-based
model of the experiment that fits essentially with the standard
interpretation of the experiment. The $CM$ system is assumed to
be monitored by the ``relative particles ''($R$) subsystem of the
atomic nucleus. The possible occurrence of decoherence is due to
the (electrons-system-mediated) interaction between $CM$ and $R$.
The interaction scales approximately as $Z^2$ ($Z$ being the
number of protons) for larger atoms providing the
$CM$-wave-packets as the approximate pointer basis.  In the case
of e.g. experimental disproving of this model, there seems only
one alternative to remain: the screen is responsible for
``reduction'' (``collapse'') of the entangled state of $CM+S$ system
($S$ denoting the atomic spin) thus mimicking existence (as
assumed in the standard interpretation of the experiment) of the
definite trajectories of $CM$ system in front of the screen.

\bigskip

{\bf Acknowledgement}: The work on this paper is financially
supported by the Ministry of Science, Serbia, under contract no
141016.

\bigskip

{\bf References}

\bigskip

\begin{flushleft}
[1] D. Bohm, ``Quantum theory'', Prentice Hall,Inc.,  New York, 1951

[2] A. Messiah, ``Quantum Mechanics'', North--Holand Publishing
    Company, Amsterdam, 1976

[3] J. von Neumann, ``Mathematical Foundations of Quantum
Mechanics'',
    Princeton University Press, Princeton, 1955

[4] T. R. Oliveira, A. O. Caldeira, ``Coherence and Entanglement
    in a Stern-Gerlach experiment'', eprint arXiv quant-ph/0608192v1
    24 Aug 2006

[5] M. Dugi\' c, Europ. Phys. J, D {\bf 29}, 173 (2004)

[6] D. Giulini, E. Joos, C. Kiefer, J. Kupsch, I.-O. Stamatescu
    and H.D. Zeh, ``Decoherence and the Appearance of a Classical
    World in Quantum Theory'', Springer, Berlin, 1996

[7] M.. Dugi\' c, Physica Scripta {\bf 53}, 9 (1996)

[8] M. Dugi\' c, Physica Scripta {\bf 56}, 560 (1997)

[9] B. L .Cohen, ``Concepts of Nuclear Physics'', McGrawHill Book
Company, New York, 1971

[10] R. E. Pierls, J. Yoccoz, ``The collective model of nuclear motion'',
in Proc. Phys. Soc. Vol. XX, p. 381, 1957

[11] M. A. Preston, R. K. Bhaduri, ``Structure of the Nucleus'',
Addison-Wesley Publ. Comp., Inc., Reading, Massachusetts, 1975

[12] H. D. Zeh, ``Roots and Fruits of Decoherence'',  Eprint arXiv:
quant-ph/0512078 v1 10 Dec 2005

[13] P. Atkins, R. Friedman, ``Molecular Quantum Mechanics'',
Oxford Univ. Press, Oxford, 2005

[14] L. A. Gribov, S. P. Mushtakova, ``Quantum Chemistry'', Gardariki, Moscow,
1999 (in Russian)

[15] J. P. Paz,  S. Habib, W. H. Zurek, Phys. Rev. D {\bf  47}, 488 (1993)

[16] A. F. Nikiforov, V. B. Uvarov ``Special Functions of Mathematical
Physics'', Birkh$\ddot{a}$user Verlag,  Basel, 1988

[17] R. Omn$\grave{e}$s, ``The Interpretation of Quantum
Mechanics'',
     Princeton University Press, Princeton, 1994

[18]Y. Hasegawa, R. Loidli, G. Badurek, M. Baron, H. Rauch,
Nature {\bf 425}, 45(2003)

[19] H. Rauch, W. Samuel, ``Neutron Interferometry'', Oxford Univ.
Press, Oxford, UK, 2000

[20] L. Hackerm$\ddot{u}$ller, S. Uttenthaler, K. Hornberger, et al, Phys.
Rev. Lett. {\bf 91}, 090408 (2003)

[21] L. Hackerm$\ddot{u}$ller, K. Hornberger, B. Brezger, et al, Nature
{\bf 427}, 711 (2004)

[22] M. Dugi\' c, J. Jekni\' c, Int. J. Theor. Phys. {\bf 45},
2215 (2006)

[23] M. Dugi\' c, J. Jekni\' c-Dugi\' c, Int. J. Theor. Physics
{\bf 47}, 805 (2008)

[24] J. Jekni\' c-Dugi\' c, M. Dugi\' c, Chin. Phys. Lett. {\bf
25}, 371  (2008)

[25] R. McWeeny, ``Methods of Molecular Quantum Mechanics'',
     Academic Press, New York, 1978

[26] M. Weissbluth, ``Atoms and Molecules'',
     Academic Press, New York, 1978

[27] I. S. Gradshteyn,  I. M. Ryzhik, ``Table of Integrals, Series,
and Products'', Academic Press, New York, 2007

\end{flushleft}

\bigskip

\noindent{\bf Appendix 1}

\bigskip

\noindent The canonical transformations eqs. (3), (4) define the
center-of-mass system $CM$ with the total mass $M$, and the set of the
``relative particles'' defined by the relative coordinate
$\hat{\vec{\rho}}_{R\alpha}$, with the reduced masses
$\mu_\alpha$ which define the kinetic terms
$\hat{T}_{R\alpha}=\displaystyle\frac{\hat{\vec{P}}^{2}_{R\alpha}}{2\mu_{\alpha}}$,
that appear in eq. (9).

Due to the distance-dependence of all of the Coulomb- and the
nuclear-interaction-terms, all these interactions obtain the form
$V(\hat{\vec{\rho}}_{R_{\alpha}})$. I.e. the {\it interactions} become
the (effective) {\it external potentials} for $R$. E.g.,
\begin{equation}
\hat{V}^{(R)}_{Coul}=k\sum_{i=1}^{Z}\sum_{j=1}^{Z}\frac{1}{|\hat{\vec{\rho}}_{R}\!\!\!^{\scriptstyle{ij}}|},
\end{equation}
where
$\hat{\vec{\rho}}_{R}\!\!\!^{\scriptstyle{ij}}=\hat{\vec{r}}_{Ei}-\hat{\vec{r}}_{pj}$,
$|k|=\displaystyle\frac{e^{2}}{4\pi\varepsilon_0}$ and, according
to eq. (7)
\begin{equation}
\hat{V}^{nn'}_{nucl}=-\gamma^{2}\frac{\exp(-\mu|\hat{\vec{\rho}}_{R}\!\!\!^{\scriptstyle{nn'}}|)}{|\hat{\vec{\rho}}_{R}\!\!\!^{\scriptstyle{nn'}}|}.
\end{equation}

It can be easily shown that,  regarding eqs. (3), (4), the
Hamiltonian obtains the form (8), where (cf. eq. (9))  appears
the internal interaction in the $R$ system [25]:
\begin{equation}
\hat{M}^{(R)}_{\eta\nu}\equiv\displaystyle
\sum_{\eta=1}^{A+Z-1}\sum_{\nu=1}^{A+Z-1}\frac{m_{\eta+1}m_{\nu+1}
{\hat{\dot{\vec{\rho}}}_{\eta}}\cdot
\hat{\dot{\vec{\rho}}}_{\nu}}{M}.
\end{equation}

Considering the {\it atom as a whole}, the $CM$ mass
$M=Z(m_e+m_p)+(A-Z)m_n$, where A
 is the  mass number of the atom. The relative masses $\mu_\alpha$ take the general form
\begin{equation}
\mu_{\alpha}=\frac{m_{\alpha+1}(M-m_{\alpha+1})}{M},
\end{equation}
where $m_{\alpha}$ are the constituent-particles  masses.

On the other side, for the {\it nucleus} center-of-mass,
$M=Zm_p+(A-Z)m_n\approx Am$, while eq. (29) gives:
\begin{equation}
\mu=(1-\frac{1}{A})m ,
\end{equation}
 where we simplify $m_n=m_p\stackrel{\sim}{=}m$.

Regarding the $CM$ and $R$ as the {\it nucleus} subsystems, the original Coulomb
interaction (cf. eq. (6)) that reads:
\begin{equation}
\hat{V}^{ep}_{Coul}=
k\displaystyle\sum_{i=1}^Z\displaystyle\sum_{j=1}^Z\displaystyle\frac{1}
{\vert\hat{\vec{r}}_{Ei}-\hat{\vec{r}}_{pj}\vert},
\end{equation}
obtains the following form due to eq. (4) :
\begin{equation}
\hat{V}^{ep}_{Coul}\equiv\hat{H}_{E+CM+R}=
k{\displaystyle\sum_{i=1}^Z}{\displaystyle\sum_{j=1}^Z}\displaystyle\frac{1}{|\hat{\vec{r}}_{Ei}-\hat{\vec{R}}_{CM}-\displaystyle\sum_{\alpha=1}^{A-1}\omega^{(j)}_{\alpha}{\hat{\vec{\rho}}}_{R\alpha}\!\!\!^{\scriptstyle{(j)}}|}.
\end{equation}
The rhs of eq. (32) is apparently a tripartite interaction
$\hat{H}_{E+CM+R}$, coupling  $E$, $CM$ and $R$, as  it appears
in eq. (12).

\bigskip

\noindent{\bf Appendix 2}

\bigskip

\noindent The kinetic terms $\hat{T}_{CM}$ and
$\hat{T}_{R_\alpha}$, as well as $\hat{T}_E$ are proportional to
$M^{-1}$, $\mu_\alpha^{-1}$, $m_e^{-1}$ respectively, where (cf.
Appendix 1) $M$ is the  total mass of the nucleus, $\mu$  is the
reduced mass for the nucleus--system and $m_e$ is the mass of an
electron.

Then  there appear the three parameters,
\begin{equation}
\kappa_1\equiv\frac{m_e}{M},\qquad
\kappa^{\alpha}_2\equiv\frac{m_e}{\mu_\alpha},\qquad
\kappa_3\equiv\frac{\mu}{M},
\end{equation}
where $\mu=min\{\mu_\alpha\}$; in eq. (30), we approximated
$\mu_{\alpha}\approx\mu$, $\forall\alpha$. So, for the {\it
realistic atoms}, $Z\stackrel{<}{\sim}10^{2}$, one may state the
following estimates:
\begin{equation}
\kappa_1\stackrel{<}\sim {10^{-4}},\qquad
\kappa_2\stackrel{<}\sim {10^{-3}},\qquad
\kappa_3\stackrel{>}\sim {10^{-2}},
\end{equation}
and $\kappa_2=max\{\kappa^{\alpha}_2\}$.

The values eq. (34) justify the applicability of the {\it
adiabatic approximation} [2, 13, 14] for $E+CM+R$ as follows: the
small values of $\kappa_{1,2}$ justify the adiabatic cut of the
electronic system $(E)$ from both $CM$ and $R$ systems, while
$CM$ and $R$ can not be properly mutually separated.

Now, the standard adiabatic approximation stems [2, 13, 14]: (a)
the exact state of $E+CM+R+S$ system reads:
\begin{equation}
\vert \chi \rangle_E \vert \Phi\rangle_{CM+R+S} + \vert
O(\kappa)\rangle_{E+CM+R+S},
\end{equation}
where $\kappa=max\{\kappa_1,\kappa_2\}$, while (b) the ``slow''
system $CM+R+S$ is described by the following effective
Hamiltonian :
\begin{equation}
\hat{H}_{CM+R+S}\cong_E\langle\chi|\hat{H}|\chi\rangle_E.
\end{equation}

For the larger systems (e.g. macromolecules, or the macroscopic
systems), due to the large number of the particles, the total
mass $M$ increases,  while the relative mass
$\mu^{'}=max\{\mu_{\alpha}\}$ does not significantly change. Then
the parameter $\kappa_3^{'}\equiv\mu^{'}/M$ may satisfy the
applicability of the adiabatic approximation also for $CM$ and
$R$ systems.

\bigskip

\noindent {\bf Appendix 3}
\bigskip

\noindent The calculation of expression (22) can be performed on
the basis of the ``addition theorem'' for the spherical harmonics [16]:

\begin{equation}
\frac{1}{|\vec{\xi}-\vec{\tau}|}=\displaystyle\sum_{s=0}^{\infty}\frac{|\rho_{<}|^{s}}{|\rho_{>}|^{s+1}}\bigg[\frac{4\pi}{2s+1}\displaystyle\sum_{m_{s}=-s}^{s}Y_s^{m_s*}(\vartheta_{\tau},\varphi_{\tau})Y_s^{m_s}(\vartheta_{\xi},\varphi_{\xi})\bigg]
\end{equation}

\noindent where $\vec{\tau}=(\tau, \vartheta_{\tau},
\varphi_{\tau})$ and $\vec{\xi}=(\xi,
\vartheta_{\xi},\varphi_{\xi})$, while $\rho_{<}=\xi$,
$\rho_{>}=\tau$ if $\xi<\tau$, and $\rho_{<}=\tau$,
$\rho_{>}=\xi$ if $\tau<\xi$. By ``$Y$'' we denote the spherical
harmonics.

Taking for the one-electron states the standard stationary states
for the hydrogen-like atom:

\begin{equation}
\phi_i(\vec{\xi})\equiv\phi_{n\ell
m}(\vec{\xi})=R_{n\ell}(\xi)Y_{\ell}^{m_{\ell}}(\vartheta_{\xi},\varphi_{\xi}),
\end{equation}

\noindent while bearing in mind $\vec{\tau} \equiv
\hat{\vec{\Omega}}$, the rhs of eq. (22) takes the following form:

\begin{eqnarray}
&\nonumber&kZ\sum_{n}\sum_{\ell=0}^{n-1}\sum_{m_{\ell}=-\ell}^{\ell}\sum_{s=0}^{\infty}\frac{4\pi}{2s+1}\sum_{m_{s}=-s}^{s}Y_s^{m_s*}(\vartheta_{\tau},\varphi_{\tau})\times
\nonumber \\&&
\times\int\frac{|\rho_{<}|^{s}}{|\rho_{>}|^{s+1}}R_{n\ell}^{2}(\xi)\xi^{2}d\xi\times
\nonumber \\&& \times\int
Y_{\ell}^{m_{\ell}}(\vartheta_{\xi},\varphi_{\xi})Y_{\ell}^{m_{\ell}*}(\vartheta_{\xi},\varphi_{\xi})Y_{s}^{m_s}(\vartheta_{\xi},\varphi_{\xi})\sin{\vartheta_{\xi}}d\varphi_{\xi}d\vartheta_{\xi}.
\end{eqnarray}

In eq. (39) we first calculate the sum
$\displaystyle\sum_{m_{\ell}=-\ell}^{\ell}I_{\vartheta\varphi}^{(m_{\ell})}$,
where $I_{\vartheta\varphi}$ is the integral over
$\vartheta_{\xi}$ and $\varphi_{\xi}$. Following the result given
in [26] (cf. eqs. (1.2-26), (1.2-27) therein), we obtain for the atoms
 with the ``closed shells'':

\begin{equation}
\sum_{m_{\ell}=-\ell}^{\ell}I_{\vartheta\varphi}^{(m_{\ell})}=\frac{2\ell+1}{\sqrt{4\pi}}\delta_{s0}\delta_{m_s0},
\end{equation}
where appear the two Kronecker-deltas.

The integration over $\xi$ in eq. (39) is performed in a few
steps. First, the integral over $\xi$ can be written as [16]:

\begin{equation}
I_{R}=\int_{0}^{\tau}\frac{\xi^{s}}{\tau^{s+1}}R_{n\ell}^{2}(\xi)\xi^{2}d\xi+\int_{\tau}^{\infty}\frac{\tau^{s}}{{\xi}^{s+1}}R_{n\ell}^{2}(\xi)\xi^{2}d\xi.
\end{equation}

Then the substitution of eqs. (40), (41) in eq. (39) gives the
following expression for eq. (39):

\begin{equation}
kZ\sum_{n}\sum_{\ell=0}^{n-1}(2\ell+1)\bigg\{\tau^{-1}\int_{0}^{\tau}R_{n\ell}^{2}(\xi)\xi^{2}d\xi+\int_{\tau}^{\infty}R_{n\ell}^{2}(\xi)\xi
d\xi\bigg\}.
\end{equation}
The integrals appearing in eq. (42) can be managed by the use of
the well-known ``incomplete'' gamma functions as defined by the
following expressions [27]:

\begin{equation}
\gamma(1+n,x)\equiv\int_{0}^{x}e^{-t}t^{n}dt=n!\bigg[1-e^{-x}\bigg(\sum_{m=0}^{n}\frac{x^{m}}{m!}\bigg)\bigg]
\end{equation}
and

\begin{equation}
\Gamma(1+n,x)\equiv\int_{x}^{\infty}e^{-t}t^{n}dt=n!e^{-x}\sum_{m=0}^{n}\frac{x^{m}}{m!}.
\end{equation}

Now, the integrations over $\xi$ give finally the following form
for $\hat{H}_{CM+R}$ (where we return $\hat \Omega$ instead of
$\tau)$:

\begin{eqnarray}
\hat{H}_{CM+R}&=\nonumber&kZ\sum_{n}\sum_{\ell=0}^{n-1}\sum_{g=0}^{n-\ell-1}\sum_{t=0}^{2g}\frac{(2\ell+1)}{2n
2^{2(n-\ell-1)}}{{2(n-\ell-1)-2g}\choose{n-\ell-1-g}}\times
\nonumber \\&&
\times\frac{(2g){!}}{g{!}(2\ell+1+g){!}}{{2g+2(2\ell+1)}\choose{2g-t}}\frac{(-2)^t}{t{!}}\nonumber\\&&\Bigg\{(2\ell+t+2){!}
\Bigg(1-\exp\bigg(-\frac{2Z\hat{\Omega}}{na_{\mu}}\bigg)\sum_{f=0}^{2\ell+t+2}\frac{(\frac{2Z\hat{\Omega}}{na_{\mu}})^{f}}{f{!}}\Bigg)\hat{\Omega}^{-1}\nonumber\\&&+
\frac{2Z}{na_{\mu}}(2\ell+t+1){!}\exp\bigg(-\frac{2Z\hat{\Omega}}{na_{\mu}}\bigg)\sum_{f=0}^{2\ell+t+1}\frac{(\frac{2Z\hat{\Omega}}{na_{\mu}})^{f}}{f{!}}\Bigg\},
\end{eqnarray}

as given in eq. (23) in the body text.

\bigskip

\end{document}